\documentclass[a4paper,11pt]{article}
\usepackage{pos}

\title{Improving $\Lambda$ Signal Extraction with Domain Adaptation via Normalizing Flows}
\ShortTitle{Improving $\Lambda$ Signal Extraction with Domain Adaptation via Normalizing Flows}

\author*[a]{Rowan Kelleher}
\author[a,1]{Matthew McEneaney}\note{Poster \ Presenter}
\author[a,b]{Anselm Vossen}

\affiliation[a]{Department of Physics, Duke University, \\ 120 Science Drive, Durham, NC 27708, USA}
\affiliation[b]{Thomas Jefferson National Accelerator Facility, \\ 12000 Jefferson Ave., Newport News, VA 23606, USA}

\emailAdd{rowan.kelleher@duke.edu}

\abstract{The present study presents a novel application for normalizing flows for domain adaptation. The study investigates the ability of flow based neural networks to improve signal extraction of $\Lambda$ Hyperons at CLAS12. Normalizing Flows can help model complex probability density functions that describe physics processes, enabling uses such as event generation. $\Lambda$ signal extraction has been improved through the use of classifier networks, but differences in simulation and data domains limit classifier performance; this study utilizes the flows for domain adaptation between Monte Carlo simulation and data. We were successful in training a flow network to transform between the latent physics space and a normal distribution. We also found that applying the flows lessened the dependence of the figure of merit on the cut on the classifier output, meaning that there was a broader range where the cut results in a similar figure of merit.}

\FullConference{25th International Spin Physics Symposium (SPIN 2023)\\
 24-29 September 2023\\
 Durham, NC, USA\\}

\begin{document}
\maketitle

\section*{Introduction}

Semi-inclusive deep inelastic scattering (SIDIS)~\cite{Spinstructure} experiments provide insight into the strong interaction that binds together the building blocks of our universe. By studying the products of scattering events we can infer information about the initial state of the struck nucleon. This paper builds on Ref.~\cite{McEneaney_2023} to improve $\Lambda$ yields in SIDIS data taken with the CLAS12 detector at Thomas Jefferson National Accelerator Facility (Jefferson Lab).

Machine learning provides novel ways to improve physics studies across the field. In the case of signal extraction, neural networks can be used to learn the similarities and differences between different scattering events that are too subtle or complex for physicists to capture with traditional studies. The $\Lambda$ Hyperon can be detected through its decay into a proton and pion: $\Lambda \Rightarrow p\pi^-$. By training a NN over a large dataset many times the model can become accurate (to ~80\% accuracy) at distinguishing events where a $\Lambda$ decayed into a proton and pion (signal) and events where a proton and pion ended up in the final state from other processes (background). However, because it is not known in measured data whether the proton and pion resulted from a $\Lambda$ decay or from a background process, we must rely on simulated events where the parents of the final state are known to train a neural network. Monte-Carlo event generation (MC)~\cite{MC_Pepsi_lund} is used to produce SIDIS events that simulate the physics that occurs in the CLAS12 detector, which is used in the present study.

Although Monte-Carlo methods work well, the simulated data does not match measured data exactly. The training data for the classification neural network differs from the data that we want to classify, and hence the performance may suffer due to the differences. Several approaches can attempt to remedy this issue, including the use of a domain-adversarial networks as used by Ref.~\cite{McEneaney_2023}. The present research aims to address the data-simulation discrepancy through the use of a relatively new class of neural networks known as normalizing flows.

\section*{Background}
\subsection*{Deep Inelastic Scattering Physics}
The CLAS12 detector at Jefferson Lab~\cite{CLAS12} enables physicists to study electron proton scattering events and gain information about the 3D structure of the nucleon. In the SIDIS process a high energy electron scatters off of a stationary target proton, and the scattered electron and at least one final state hadron are measured in the detector. The beam electron produces a virtual photon with four-momentum $q$ which is absorbed by a quark, knocking the quark out of the bound proton state. Due to color confinement, the struck quark fragments into one or more hadrons which can be measured by the detector. Hadrons can also be produced by the remaining quarks in the target proton. Measurements of the final state hadrons can inform the initial state of the proton, making them appealing candidates for study.

Relevant kinematics include: 

\begin{gather}
Q^2 = -q^2 \qquad \qquad W^2 = (P + q)^2 \qquad \qquad \nu = E - E' \\ \\
x = \frac{Q^2}{2M \nu} \qquad \qquad y = \frac{P \cdot q}{P \cdot l} \qquad \qquad z = \frac{P \cdot P_h}{P \cdot q}\\ \\
x_F = \frac{2 P_h \cdot q}{W |q|}
\end{gather}

\subsection*{Data}
This study utilizes the same data used in Ref.~\cite{McEneaney_2023}. This dataset was taken during the fall 2018 run in the outbending toroidal configuration and 10.6 GeV polarized electron beam with an unpolarized liquid hydrogen target. Each event was required to have an identified scattered electron and proton-pion pair in reconstruction. The electron was also required to be the trigger particle, have the highest momentum of all electrons in the event, and be measured in the forward detector. The PID quality was required to be $|\chi ^2| < 3$ for the scattered electron. Lambda's were selected through a series of kinematic cuts: $Q^2 > 1 GeV^2$, $W > 2 GeV$, $y < 0.8$, $z_{p\pi^-} < 1$, $x_F > 0$, as well as a cut on the invariant mass of the $p\pi^-$ pair placed at $M_{p\pi^-} < 1.24$.
\subsection*{Monte Carlo Simulation Events}
The Monte Carlo simulation was produced with the same configuration as the data from the fall 2018 run, generated by an algorithm based on the Pepsi Lund program~\cite{MC_Pepsi_lund}. Events where a $\Lambda$ decays into a $p\pi^-$ in the MC truth were tagged as signal, while $p\pi^-$ pairs without a $\Lambda$ parent were tagged as background.

\section*{Machine Learning Methods}
Normalizing Flows~\cite{Normalizing_flow_intro} have been implemented across several fields, notably in image generation and recently high energy and nuclear physics. Normalizing Flows utilize a series of bijective, computationally efficient functions to transform a distribution with an unknown probability distribution function (PDF) to a distribution with a known PDF. The PDF of the unknown distribution can then be calculated through the change of variables formula. The functions are represented by multilayer-perceptrons (MLPs) and often many layers are used together to create more complex transformations. By training the network many times, the network can learn functions that best approximate the transformation that matches the unknown distribution. Because the functions are bijective the network can be reversed: samples from a known distribution can be transformed to create new samples of the target distribution.

The present study attempts to transform data to a simulation-like distribution using normalizing flows to solve the discrepency between data and MC. By training two flow models, one on the measured data and one on the simulation, we can transform the data first to a normal distribution space by using the data-NN, then backwards through the simulation-NN to the simulation space. If done properly, this process could improve classifier performance as the inputs to the classifier would more closely match the classifier training data.

\subsection*{Model Architecture}
This study uses a realNVP architecture where coupling layers are utilized to improve computation efficiency. The realNVP architecture was introduced by Ref.~\cite{dinh2017density}  to provide efficient networks for density estimation. Because the change of variables formula depends on calculating the determinant of the transformation Jacobian, the calculation can be expensive. However, affine coupling layers address this issue through the properties of triangular matrices. Because the determinant of a triangular matrix is equal to the product of the diagonals, the computation of the determinant is efficient if we choose functions that produce a triangular Jacobian. To accomplish this, we split the input x into two components, leaving the outputs of the first component unaltered, and parameterize the outputs of the second component using the first component. Given an input $x$ of dimension $D$, and $d < D$, we can write:
$$y_{1:d} = x_{1:d}$$
$$y_{d+1:D}=x_{d+1:D}\cdot exp(s(x_{1:d})) + t(x_{1:d})$$ 
Here, s and t are scale and translation functions. When we calculate the jacobian, we can see that the upper right corner of the 2x2 matrix is 0, meaning that the matrix is lower triangular, allowing us to calculate the determinant more efficiently.~\cite{dinh2017density}

The separation of the two components is implemented using a binary bitmask b that alternates between 0s and 1s: $$y = b \cdot x + (1 - b) \cdot \left(x \cdot exp(s(b \cdot)) + t(b \cdot x) \right)$$
\section*{Normalizing Flow Study}
\subsection*{Classifier Input Transformation}
A GNN (as described in Ref.~\cite{McEneaney_2023}) was trained by maximizing classifier accuracy on simulation data, where the classifier tried to distinguish signal and background events. The layers of the GNN prior to the classifier were then disconnected and used to extract the features of the events into a latent space with a fixed dimension of 71. This latent space was then used as the input to the Normalizing Flow model. We utilized the normflows python package (Ref.~\cite{Stimper2023}) which has functionality for masked affine layers and base distributions. The NF models had an input and output dimension (which must be the same so that they are bijective) of 71. The MLPs modelling the scale and translation functions had hidden dimensions of 142, and each MLP had 2 hidden layers. We used 52 masked affine layers and alternated the bitmask in each layer so that each input dimension was transformed every other layer. The gaussian base distributions were 71 dimensional to match the inputs, and had means of 0 and standard deviations of 1. 

The training procedure consisted of sampling a batch of 100 events from the GNN output dataset, transforming the batch using the current state of the NF model, then calculating the log-probability that the NF output was drawn from the base distribution. The model was then updated using gradient descent. We also utilized the validation dataset to calculate validation loss, but did not update the model according to this loss. This process is repeated for every batch, and then again for numerous epochs. We trained the MC model over 11 epochs and the DATA model over 6 epochs (the data model was trained for fewer epochs because the dataset was larger, and hence the loss plateaued after fewer epochs). After training we used the reserved testing dataset to calculate an average loss to ensure that the models did not overfit or diverge.

\begin{figure}[h!]
    \centering
    \includegraphics[width=0.45\textwidth]{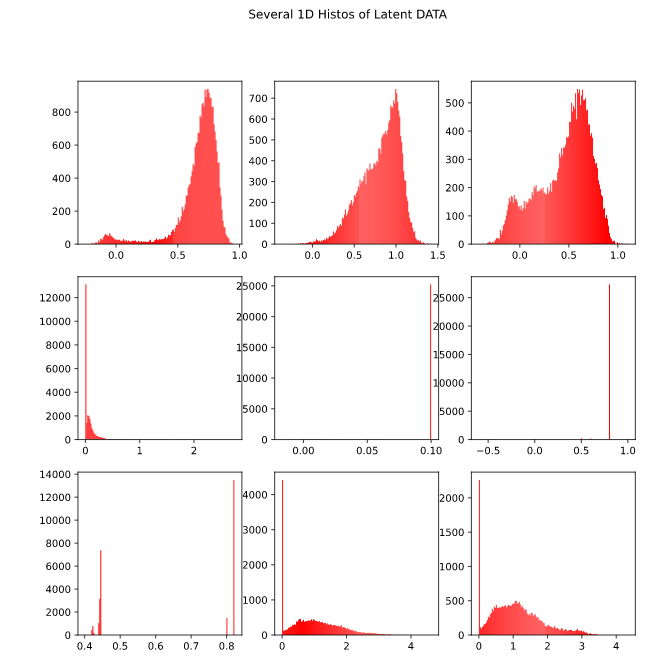}
    \includegraphics[width=0.45\textwidth]{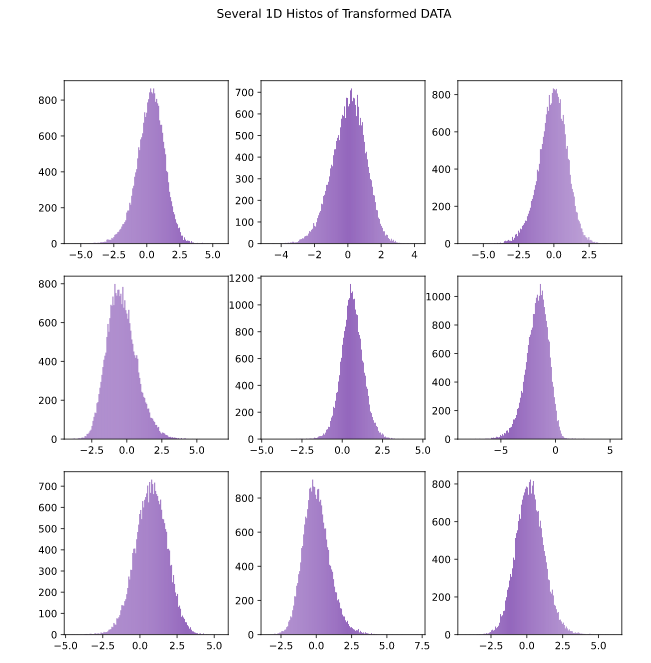}
    \caption{A few dimensions are shown for the latent data (left) and the normalized data (passed through the first NF model, but not the second) (right).}
    \label{fig:many_histos}
\end{figure}

\subsection*{Distortion Reversal}
In addition to the study on improving classifier performance, we also studied the ability for normalizing flows to reverse distortions found in data. In this study we used three kinematic variables of the proton and pion instead of the latent representation to investigate how well the neural networks could recover an underlying distribution after being distorted. The goal was to start with some simulation data, add a distortion value to each event, and then transform the distorted data back to the simulation data. We trained two different NF models, one for the MC kinematics, and one for the distorted kinematics. We chose to use transverse momentum $p_T$, azimuthal angle $\phi$, and polar angle $\theta$ for both the proton and pion in every event as the NN input. We distorted only the proton $p_T$, leaving the other 5 inputs as their MC values. Two types of distortion were studied: sampling distortion values for each event from a gaussian distribution; and adding the same value to every event to shift the proton pT distribution. The standard deviation of the gaussian distribution was varied to study different "distortion intensities," and the shift value was varied for shift distortion. The mean of the gaussian was kept constant at 0.

\section*{Results}
\subsection*{Classifier Input Performance}
\begin{figure}[t]
  \centering
  \includegraphics[width=0.45\textwidth]{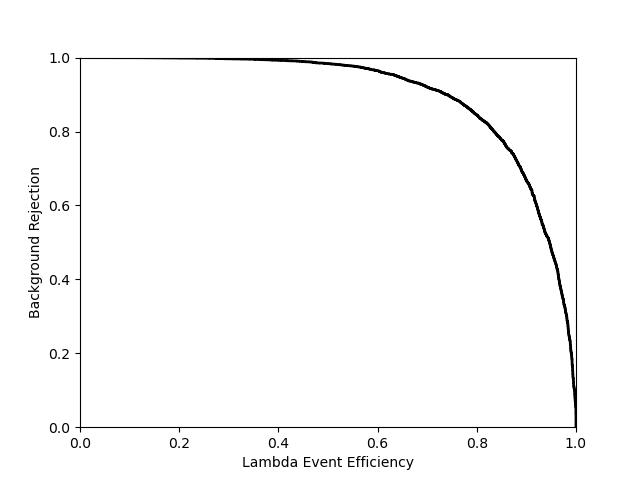}
  \includegraphics[width=0.45\textwidth]{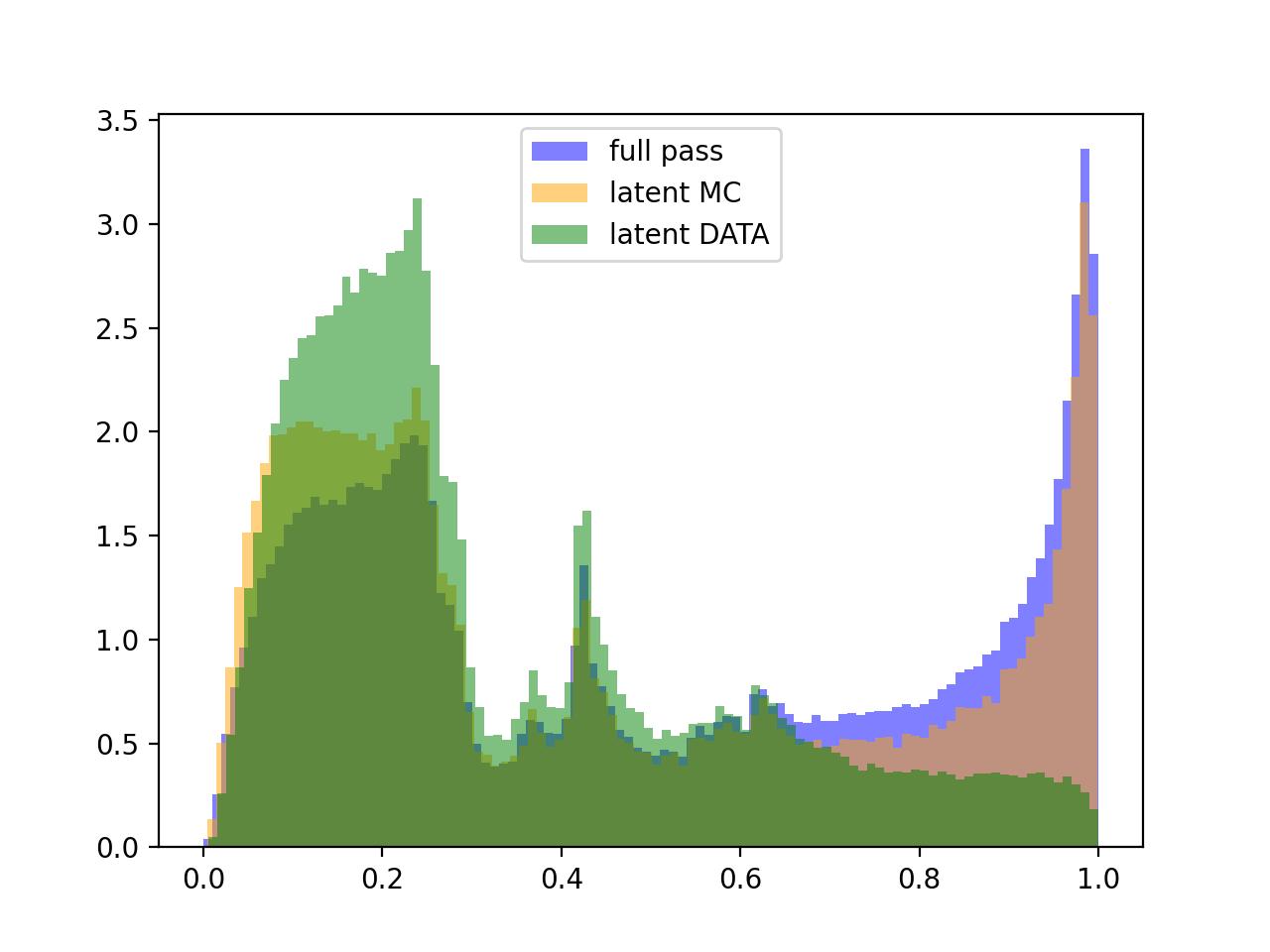}
  \caption{ROC curve (left) for the latent MC and classifier output across latent data, transformed data, and MC. The ROC curve had an AUC of 0.90}
  \label{fig:roc_curve_classifier_output}
\end{figure}

The signal peak for the data mass distribution was fit using a Crystal Ball signal function~\cite{CrystalBall} and a quadratic background. We evaluated the normalizing flow performance through the use of the figure of merit (FOM = $N_{signal} / \sqrt{N_{tot}}$) on the signal peak between $\pm 2 \sigma$ around the fitted signal peak $\mu$. We used 100 bins for the mass spectrum and integrated over the bins falling in the peak region to calculate $N_{tot}$. $N_{signal}$ was calculated by fitting the mass spectrum's bins to the the signal function, and then integrating over the fit histogram. The figure of merit and purity were calculated for 15 different cuts on the roc curve (shown in figure \ref{fig:roc_curve_classifier_output}) to illustrate performance as a function of the cut. 
\begin{figure}[t]
  \centering
  \includegraphics[width=0.45\textwidth]{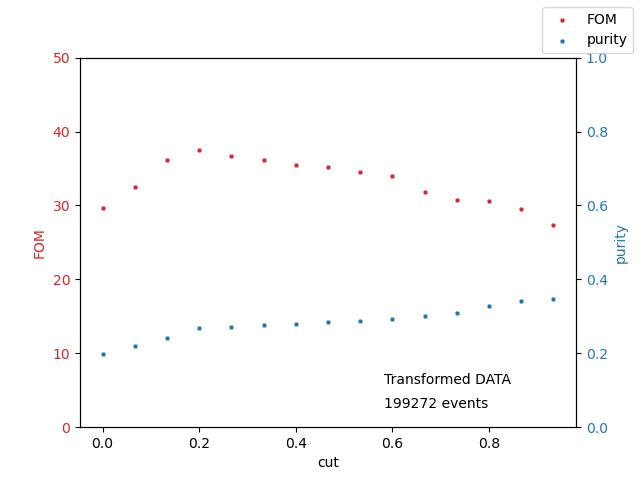}
  \includegraphics[width=0.45\textwidth]{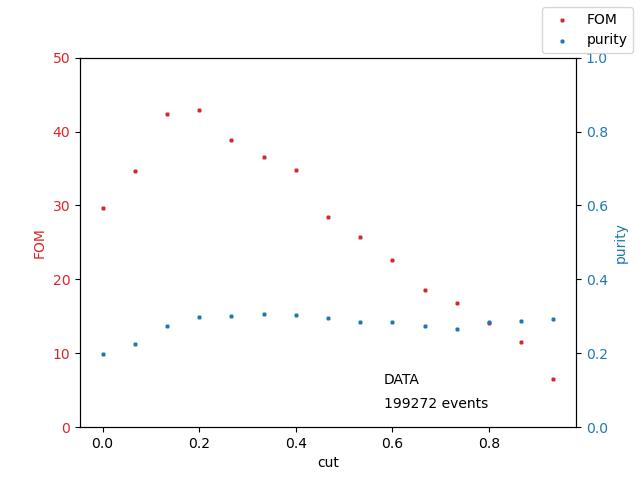}
  \caption{FOM and purity for transformed data (left) and non-transformed data (right).}
  \label{fig:FOM_plots}
\end{figure}

Figure \ref{fig:roc_curve_classifier_output} shows how the classifier output of the transformed data matches that of the MC while the classifier output of the latent data does not, specifically at larger cut values. As shown in figure \ref{fig:FOM_plots}, the figure of merit calculated for the latent data falls sharply for higher cuts on the classifier output where the classifier output of the latent data fails to match that of MC. The figure of merit of the transformed data stays much flatter than that of the latent data. Having a flatter figure of merit can help improve generalizability as the performance of the classifier will perform similarly at all cut values, meaning there is a smaller dependency on classifier cut. 

\subsection*{Distortion Performance}
\begin{figure}[h!]
  \centering
    \includegraphics[width=0.45\textwidth]{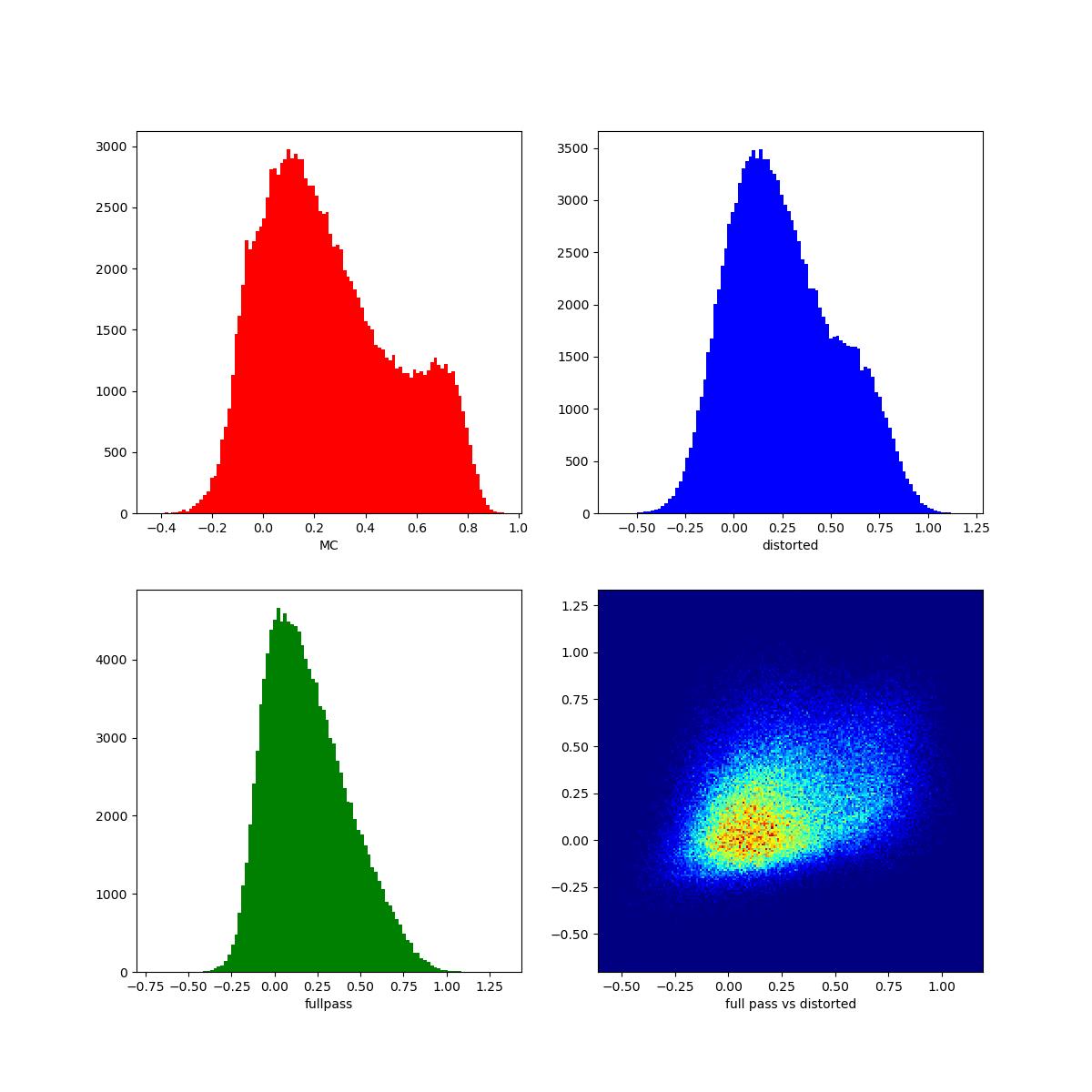}
    \includegraphics[width=0.45\textwidth]{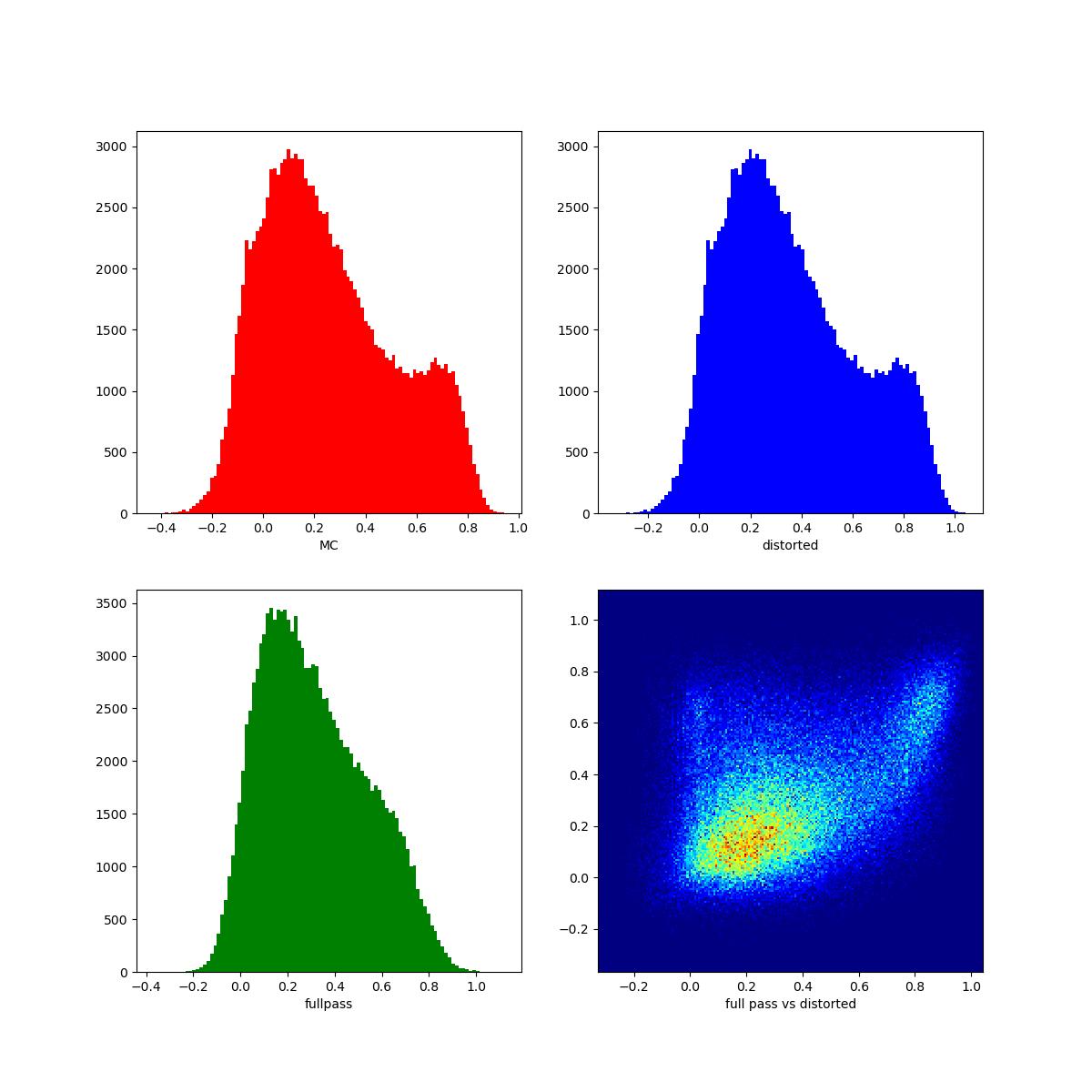}
  \caption{Plots of the proton $p_T$ distribution for the MC $p_T$, distorted $p_T$, and transformed $p_T$ with transformed $p_T$ plotted against distorted $p_T$. Proton $p_T$ is distorted by adding a value drawn from a normal distribution with $\sigma = 0.1$ (left) and by shifting every $p_T$ 0.1 GeV (right)}
  \label{fig:Distortion}
\end{figure}

The original Monte Carlo, distorted, and NF-output distributions are plotted in figure \ref{fig:Distortion}, showing the transformation of the distribution across the process. Smearing matrices (in this case transformed $p_T$ plotted against distorted $p_T$) were used to observe the effectiveness of this approach in restoring the distorted distribution. 

The normalizing flow transformation did not successfully recover the original MC distribution, but instead recovered broader features of the original distribution while losing finer features. The distorted distribution for a distortion of 0.1 GeV is shown in figure \ref{fig:Distortion} for both the gaussian and shift distortions. In the shift distortion, the transformed distribution peaks at a closer value to the MC, meaning the transformation was able to shift the distribution in the correct direction. However, the distribution loses its features, such as the secondary peak around 0.7 GeV in the MC. The transformation process appears to turn the distribution closer to a gaussian distribution. Similarly, the gaussian-distorted distribution becomes much more normal when transformed.

Another consideration with this approach is the effect that the process has on the non-distorted dimensions. Because the NF models rely on fully connected networks, the dimensions that do not need restoring are affected by the transformation, meaning that the process must learn to restore the distorted dimension while leaving the other dimensions in similar shapes. In our study, the models struggled at this task, often turning more complex distributions into single peaks.

To address the issue of the NF altering non-distorted dimensions, conditional NF models could be used, where the input is only the distorted dimension, and the conditions are the other dimensions. Furthermore, using different base distributions, such as multi-modal normal distributions, could help solve these issues, and are potential areas of further study.
\section*{Conclusion}
Normalizing Flows have many possible use cases in nuclear physics that have not been investigated yet. This study attempted two use cases: transforming latent representations of SIDIS data to an MC-like latent space to improve particle identification; restoring distorted kinematic distributions to their expected state. This study has shown that there are promising ways to utilize flows to improve $\Lambda$ identification at CLAS12, and suggests future subjects of study to improve the present methods.

\bibliographystyle{JHEP}
\bibliography{proceedings.bib}

\end{document}